\documentstyle[emulateapj]{article}

\lefthead{Nagao et al.}
\righthead{Iron is not Depleted in HINERs}

\begin{document}

\submitted{The Astronomical Journal, in Press} 
\title{Iron is not Depleted in High-Ionization Nuclear Emission-Line 
       Regions of Active Galactic Nuclei}

\author{Tohru NAGAO, Takashi MURAYAMA, Yasuhiro SHIOYA, 
        and Yoshiaki TANIGUCHI}
\affil{Astronomical Institute, Graduate School of Science, 
       Tohoku University, Aramaki, Aoba, Sendai 980-8578, Japan\\
       tohru@astr.tohoku.ac.jp, murayama@astr.tohoku.ac.jp, 
       shioya@astr.tohoku.ac.jp, tani@astr.tohoku.ac.jp}


\begin{abstract}

In order to examine whether or not high-ionization nuclear emission-line 
regions (HINERs) in narrow-line regions of active galactic nuclei 
are dusty, we focus on two high-ionization forbidden emission
lines, [Fe {\sc vii}]$\lambda$6087 and [Ne {\sc v}]$\lambda$3426.
We perform photoionization model calculations to investigate possible
dependences of the flux ratio of 
[Fe {\sc vii}]$\lambda$6087/[Ne {\sc v}]$\lambda$3426 
on various gas properties, in order to investigate
how useful this flux ratio to explore the dust abundances in HINERs.
Based on our photoionization model calculations, we show that the 
observed range of the flux ratio of 
[Fe {\sc vii}]$\lambda$6087/[Ne {\sc v}]$\lambda$3426 is consistent with
the dust-free models while that is hard to be explained by the 
dusty models. This suggests that iron is not depleted at HINERs, which
implies that the HINERs are not dusty.
This results is consistent with the idea that the HINERs are located
closer than the dust-sublimation radius (i.e., inner radius of dusty tori)
and thus can be hidden by dusty tori when seen from a edge-on view 
toward the tori, which has been also suggested by the AGN-type 
dependence of the visibility of high-ionization emission lines.

\end{abstract}



\section{INTRODUCTION}

The narrow-line region (NLR) is one of the fundamental ingredients of
active galactic nuclei (AGNs) such as Seyfert galaxies, and thus its 
physical and chemical properties have
been studied intensively up to now (see Osterbrock \& Mathews 1986 for
a review). Since the presence of dust grains significantly 
affects emission-line spectra in several ways, it has been often discussed
whether or not dust grains survive in the NLRs. 
The chemical composition of the gas phase is modified by the depletion
of refractory elements into the grains. Electrons photoelectrically
ejected from the grains heat the gas, while electron-grain collisions cool
it. The dust opacity modifies the transfer of the ionizing continuum
that irradiates gas clouds in NLRs. 
Therefore the knowledge of dust abundances is required to interpret
observed emission-line spectra of NLRs (see, e.g., Ferguson et al. 1997a).

There are some pieces of evidence that suggest the presence of dust
grains in NLRs.
Asymmetric narrow emission-line profiles may suggest the presence of
dust grains in the gas (e.g., Heckman et al. 1981; De Robertis \&
Osterbrock 1984; Whittle 1985a, 1985b).
In partially-ionized regions in NLRs the depletion of some refractory
elements is suggested by the weakness of the [Ca {\sc ii}] emission
(e.g., Kingdon, Ferland, \& Feibelman 1995; Ferguson et al. 1997a)
and by comparing the intensity of near-infrared [Fe {\sc ii}] emission
lines with that of [O {\sc i}]$\lambda$6300 (e.g., Mouri et al. 1989;
Simpson et al. 1996; Alonso-Herrero et al. 1997; 
Mouri, Kawara, \& Taniguchi 2000) and with
that of [P {\sc ii}]1.188$\mu$m (e.g., Oliva et al. 2001; 
Rodr\'{\i}guez-Ardila et al. 2002).
Some resonance lines have been observed to be very weak, which seems to be also
due to line transfer effect within dusty gas (e.g., Kraemer et al. 2000).
These naturally lead to the following question;
is the NLR ubiquitously dusty? 

The innermost region of NLRs, which is irradiated by high photoionizing 
flux, is thought to be rather high-velocity,
highly-ionized, dense gas clouds (e.g., De Robertis \& Osterbrock 1986;
Murayama \& Taniguchi 1998a, 1998b; Barth et al. 1999;
Tran, Cohen, \& Villar-Martin 2000; Nagao, Taniguchi, \& Murayama 2000b;
Nagao, Murayama, \& Taniguchi 2001a, 2001b).
Since the gas clouds in this region can radiate very high ionization
emission lines such as [Ne {\sc v}], [Fe {\sc vii}], and [Fe {\sc x}] 
(the so-called coronal lines), such regions are called ``coronal
line regions'' or ``high-ionization nuclear emission-line regions''
(HINERs: Binette 1985; Murayama, Taniguchi, \& Iwasawa 1998).
Nussbaumer \& Osterbrock (1970) reported that the gas-phase abundance
of iron that is estimated by using the emission-line flux
ratio of [Fe {\sc vii}]$\lambda$6087/[Ne {\sc v}]$\lambda$3426
is close to the solar value. This means that the iron is not depleted
significantly because of the absence of grains in HINERs. 
However, Ferguson, Korista, \& Ferland (1997b) mentioned that
the flux ratio of [Fe {\sc vii}]$\lambda$6087/[Ne {\sc v}]$\lambda$3426
depends on physical properties of gas clouds and thus it may be
inappropriate to use this ratio as an indicator of 
the gas-phase iron abundance.
They proposed alternative indicators of the gas-phase abundances of
refractory elements by using some near-infrared high-ionization lines
such as [Ca {\sc viii}]2.32$\mu$m, and then suggested that 
the HINERs seem to be dust-free (see also Ferguson et al. 1997a).
Although their method is a robust one, it can be applied only to 
very few objects since the near-infrared high-ionization lines are
hard to be measured. Indeed they applied their method only on the 
two Seyfert galaxies, NGC 1068 and the Circinus galaxy,
and they concluded that the HINERs in the two Seyfert galaxies
are dust free.

\begin{figure*}
\epsscale{0.80}
\plotone{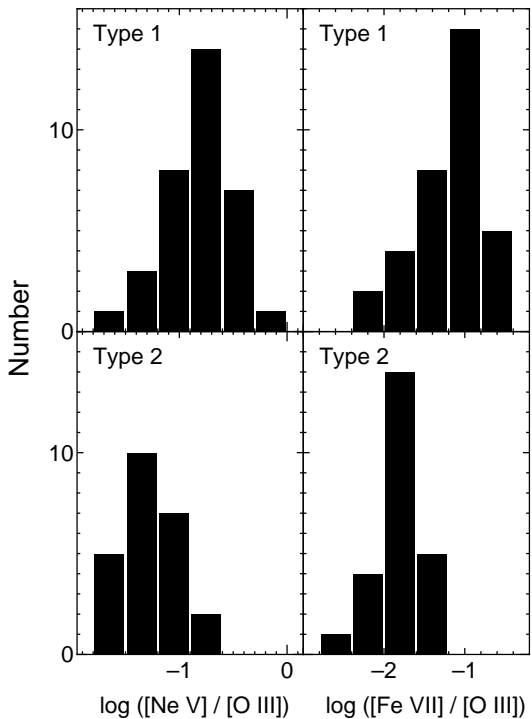}
\caption{
Frequency distributions of the two emission-line flux
ratios, [Ne {\sc v}]$\lambda$3426/[O {\sc iii}]$\lambda$5007 and
[Fe {\sc vii}]$\lambda$6087/[O {\sc iii}]$\lambda$5007,
for the type 1 AGNs and the type 2 AGNs in our sample.
\label{fig1}}
\end{figure*}

In order to investigate whether the HINERs contain dust grains 
for a large sample of AGNs, we focus on 
[Fe {\sc vii}]$\lambda$6087 and [Ne {\sc v}]$\lambda$3426 again.
These two forbidden emission lines are thought to arise at similar regions
since their critical densities are similar
($1.6 \times 10^7$ cm$^{-3}$ and $3.6 \times 10^7$ cm$^{-3}$, respectively),
and the ionization potentials of Fe$^{6+}$ and Ne$^{4+}$ are also nearly the 
same (97.1 eV and 99.1 eV, respectively). 
In addition to this advantage, it should be also an great advantage that
these two emission lines are enough strong to be measured easily.
Therefore, they are useful to examine the dust abundances
in the HINERs if we know some possible dependences of their emissivities
on physical properties, even though they might be inappropriate 
to determine exact gas-phase elemental abundances as noted 
by Ferguson et al. (1997b).
In this paper, we investigate the dependences of the two high-ionization
emission lines on physical properties based on photoionization model
calculations. We then examine how these emission lines can give constraints
on the issue whether or not the HINERs are dusty.

\section{DATA OF EMISSION-LINE FLUX RATIOS}

To study the properties of gas clouds in HINERs, we have compiled 
the data of emission-line flux ratios of
[Fe {\sc vii}]$\lambda$6087/[O {\sc iii}]$\lambda$5007 and
[Ne {\sc v}]$\lambda$3426/[O {\sc iii}]$\lambda$5007 from the literature.
The details of the data compilation are given by Nagao et al. (2001b). 
The number of compiled objects is 58; 34 type 1 AGNs and 24 type 2 AGNs.
Here we refer to V\'{e}ron-Cetty \& V\'{e}ron (2001) for the AGN type of 
each object. Note that, in this paper, the objects classified as
type 1.0, 1.2, and 1.5 AGN (including narrow-line Seyfert 1 galaxies) by
V\'{e}ron-Cetty \& V\'{e}ron (2001) are included in ``type 1 AGN'',
and the objects classified as type 1.8, 1.9, and 2.0 AGN are included
in ``type 2 AGN''.
Since we do not impose any selection criteria upon our sample, 
this sample is neither an uniform nor complete one in any sense.
However, this does not affect the following discussion significantly because
we are not interested in statistical properties of NLR gas clouds.
Namely, we focus on whether distribution of emission-line flux ratios 
is consistent with that expected for dust-free cases or dusty cases.
The compiled emission-line flux ratios are given in Table 1.
We adopt an average value if an emission-line flux ratio for a 
certain object is given in more than one paper published previously.
As for the data of the emission-line flux ratios, we do not make any 
reddening correction since it is often difficult to measure the fluxes
of narrow components of Balmer lines for type 1 AGNs
(see, e.g., Nagao et al. 2001b).
Effects of the dust extinction are discussed when necessary.

\section{RESULTS}

\begin{figure*}
\epsscale{0.45}
\plotone{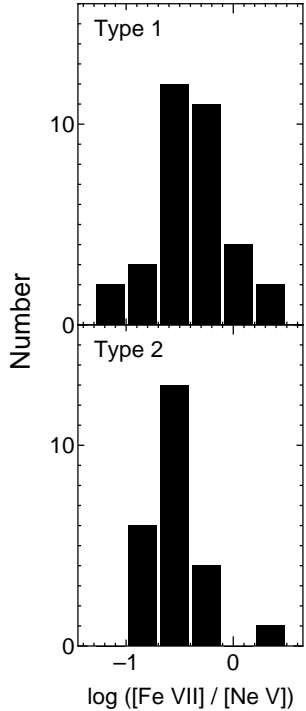}
\caption{
Frequency distributions of the emission-line flux ratio of
[Fe {\sc vii}]$\lambda$6087/[Ne {\sc v}]$\lambda$3426
for the type 1 AGNs and the type 2 AGNs in our sample.
\label{fig2}}
\end{figure*}

In Figure 1, we show the frequency distributions of the two
emission-line flux ratios, 
[Ne {\sc v}]$\lambda$3426/[O {\sc iii}]$\lambda$5007 and
[Fe {\sc vii}]$\lambda$6087/[O {\sc iii}]$\lambda$5007, for 
both the type 1 AGNs and the type 2 AGNs in our sample.
The average and the 1-$\sigma$ standard deviation of these 
two emission-line flux
ratios are 0.191 $\pm$ 0.114 and 0.089 $\pm$ 0.079 for the type 1 AGNs
and 0.066 $\pm$ 0.042 and 0.020 $\pm$ 0.015 for the type 2 AGNs, 
respectively. The comparison of these results between the type 1 and 
type 2 AGNs clearly suggests that the type 1 AGNs exhibit systematically 
larger ratios of both [Ne {\sc v}]$\lambda$3426/[O {\sc iii}]$\lambda$5007 
and [Fe {\sc vii}]$\lambda$6087/[O {\sc iii}]$\lambda$5007 than the 
type 2 AGNs. 
We apply the Kolmogorov-Smirnov (K-S) statistical test on the data to
estimate the significance of the differences in the two emission-line
flux ratios between the type 1 AGNs and the type 2 AGNs.
The resultant probabilities that the frequency distributions of
the type 1 and type 2 AGNs come from the same underlying populations
are $2.0 \times 10^{-6}$ and $5.9 \times 10^{-6}$ for
[Ne {\sc v}]$\lambda$3426/[O {\sc iii}]$\lambda$5007 and
[Fe {\sc vii}]$\lambda$6087/[O {\sc iii}]$\lambda$5007, respectively.
It is thus confirmed that the two emission-line flux ratios are 
significantly larger in the type 1 AGNs than in the type 2 AGNs.
This result is consistent with the previous reports 
by Murayama \& Taniguchi (1998a), Nagao et al. (2000b), and 
Nagao et al. (2001b). They concluded that these differences in
the emission-line flux ratios are due to the orientation effect;
i.e., the HINER is located very close to the nucleus and thus can be
hidden by dusty tori if it is observed from an edge-on view
toward the tori.

Contrary to the above results, the difference in the flux ratio of 
[Fe {\sc vii}]$\lambda$6087/[Ne {\sc v}]$\lambda$3426 between
the type 1 and type 2 AGNs is fairly small, as shown in Figure 2.
The average and the 1-$\sigma$ standard deviation of this
flux ratio are 0.523 $\pm$ 0.461 for the type 1 AGNs and
0.335 $\pm$ 0.315 for the type 2 ones. 
The average and the 1-$\sigma$ standard deviation for
the all sample (i.e., 58 AGNs) are 0.445 $\pm$ 0.415, and
the median value is 0.309.
Although the flux ratios of the type 2 AGNs seem to be somewhat
smaller than the type 2 AGNs, the difference is not significant.
The K-S statistical test results in the K-S probability of
1.1 $\times$ 10$^{-2}$, which means that the inferred difference is
marginal.

\section{PHOTOIONIZATION MODEL CALCULATIONS}

In order to examine whether or not the HINERs are dusty,
we investigate how the emission-line flux ratio of
[Fe {\sc vii}]$\lambda$6087/[Ne {\sc v}]$\lambda$3426 depends on 
some physical properties by performing calculations of photoionization 
models with and without dust grains.
Our method and results of the photoionization model calculations
are presented below.

\subsection{Method}

We carry out several photoionization model calculations by using
the publicly available code $Cloudy$ version 94.00 (Ferland 1997, 2000).
Here we assume uniform density gas clouds with a plane-parallel geometry.
The parameters for the calculations are:
(1) the spectral energy distribution (SED) of the input continuum radiation;
(2) the hydrogen density of a cloud ($n_{\rm H}$);
(3) the ionization parameter ($U$), i.e., the ratio of the ionizing
    photon density to the hydrogen density at the irradiated surface 
    of a cloud;
(4) the column density of a cloud ($N_{\rm H}$); and
(5) the elemental composition and the dust abundance of the gas.

We adopt the SED in the form of $f_{\nu} \propto \nu^{\alpha}$ with
$\alpha = 2.5$ for $\lambda > 10$ $\mu$m, $\alpha = -1.5$ between
10 $\mu$m and 50 keV, and $\alpha = -2.0$ for $h\nu > 50$ keV,
taking account of the SEDs actually observed in AGNs
(e.g., Koski 1978; Storchi-Bergmann \& Pastoriza 1989, 1990;
Kinney et al. 1991; see also Ho, Shields, \& Filippenko 1993).
Since forbidden lines arise at gas clouds with a density near their
critical densities most effectively, the gas density of the
[Fe {\sc vii}]$\lambda$6087 and the [Ne {\sc v}]$\lambda$3426 emitting 
regions are expected to be a few $\times 10^7$ cm$^{-3}$.
We thus perform model runs with $n_{\rm H} = 10^{6.5}, 10^{7.0}$, and
$10^{7.5}$ cm$^{-3}$. 
As for the ionization parameter, we investigate models with
$U = 10^{-2.5}$, $10^{-2.0}$ and $10^{-1.5}$, although the
ionization parameter of HINERs is thought to be rather high, that is
to say, $U \gtrsim 10^{-2}$ (e.g., Murayama \& Taniguchi 1998b;
Nagao et al. 2001b).
Because the the column density of a cloud may be much different
from object to object, we perform model runs in the range of
$10^{20.0}$ cm$^{-2}$ $\le$ $N_{\rm H}$ $\le$ $10^{22.0}$ cm$^{-2}$.

For the dust-free models, we set the gas-phase elemental abundances
to be the solar ones, which are taken from Grevesse \& Anders (1989)
with extensions by Grevesse \& Noels (1993). 
The adopted gas-phase elemental abundances are:
H:  1.00, 
He: 1.00$\times$10$^{-1}$,
Li: 2.04$\times$10$^{-9}$,
Be: 2.63$\times$10$^{-11}$,
B:  7.59$\times$10$^{-10}$,
C:  3.55$\times$10$^{-4}$,
N:  9.33$\times$10$^{-5}$,
O:  7.41$\times$10$^{-4}$,
F:  3.02$\times$10$^{-8}$,
Ne: 1.17$\times$10$^{-4}$,
Na: 2.06$\times$10$^{-6}$,
Mg: 3.80$\times$10$^{-5}$,
Al: 2.95$\times$10$^{-6}$,
Si: 3.55$\times$10$^{-5}$,
P:  3.73$\times$10$^{-7}$,
S:  1.62$\times$10$^{-5}$,
Cl: 1.88$\times$10$^{-7}$,
Ar: 3.98$\times$10$^{-6}$,
K:  1.35$\times$10$^{-7}$,
Ca: 2.29$\times$10$^{-6}$,
Sc: 1.58$\times$10$^{-9}$,
Ti: 1.10$\times$10$^{-7}$,
V:  1.05$\times$10$^{-8}$,
Cr: 4.84$\times$10$^{-7}$,
Mn: 3.42$\times$10$^{-7}$,
Fe: 3.24$\times$10$^{-5}$,
Co: 8.32$\times$10$^{-8}$,
Ni: 1.76$\times$10$^{-6}$,
Cu: 1.87$\times$10$^{-8}$, and
Zn: 4.52$\times$10$^{-8}$.
For the dusty models, Orion-type graphite and silicate grains
(Baldwin et al. 1991; see also Ferland 1997) are contained in a gas cloud.
As a result of the metal depletion onto dust grains, the gas-phase 
elemental abundances are altered to:
H:  1.00, 
He: 9.50$\times$10$^{-2}$,
Li: 5.40$\times$10$^{-11}$,
B:  8.90$\times$10$^{-11}$,
C:  3.00$\times$10$^{-4}$,
N:  7.00$\times$10$^{-5}$,
O:  4.00$\times$10$^{-4}$,
Ne: 6.00$\times$10$^{-5}$,
Na: 3.00$\times$10$^{-7}$,
Mg: 3.00$\times$10$^{-6}$,
Al: 2.00$\times$10$^{-7}$,
Si: 4.00$\times$10$^{-6}$,
P:  1.60$\times$10$^{-7}$,
S:  1.00$\times$10$^{-5}$,
Cl: 1.00$\times$10$^{-7}$,
Ar: 3.00$\times$10$^{-6}$,
K:  1.10$\times$10$^{-8}$,
Ca: 2.00$\times$10$^{-8}$,
Ti: 5.80$\times$10$^{-10}$,
V:  1.00$\times$10$^{-10}$,
Cr: 1.00$\times$10$^{-8}$,
Mn: 2.30$\times$10$^{-8}$,
Fe: 3.00$\times$10$^{-6}$,
Ni: 1.00$\times$10$^{-7}$,
Cu: 1.50$\times$10$^{-9}$, and
Zn: 2.00$\times$10$^{-8}$.
These depleted elemental abundances are based on results of several
studies of the Orion nebula (Baldwin et al. 1991; Rubin et al. 1991;
Rubin, Dufour, \& Walter 1992a; Rubin et al. 1992b; 
Osterbrock, Tran, \& Veulleux 1992).
Note that beryllium, fluorine, scandium and cobalt are not included in the
calculations for the dusty models.
See Ferland (1997) for details of the treatments of elemental abundances
and dust grains.

\subsection{Results of the Model Calculations}

\begin{figure*}
\epsscale{1.00}
\plotone{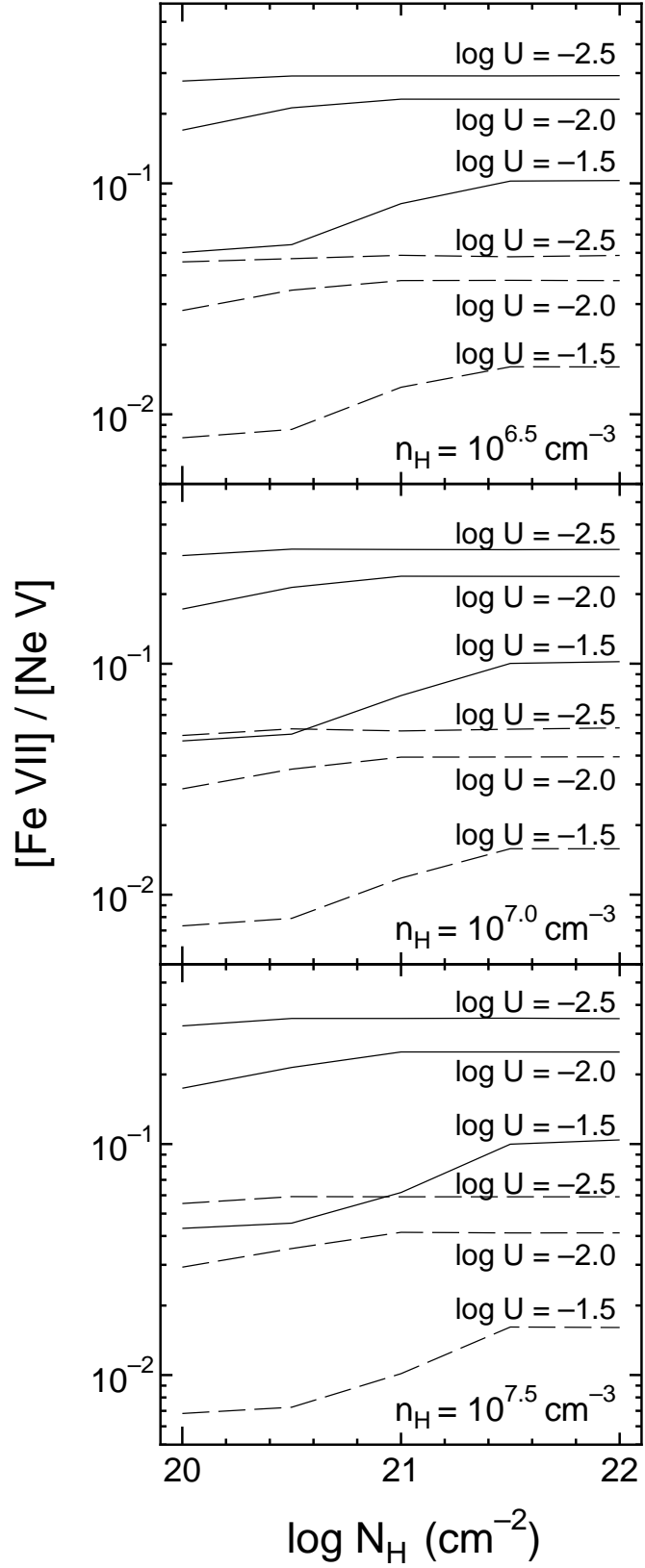}
\caption{
Predicted emission-line flux ratio of
[Fe {\sc vii}]$\lambda$6087/[Ne {\sc v}]$\lambda$3426 for the
dust-free models ($solid$ $lines$) and the dusty models ($dashed$ $lines$).
The results for the cases of $n_{\rm H} = 10^{6.5}$ cm$^{-3}$
($upper$ $panel$), $10^{7.0}$ cm$^{-3}$ ($middle$ $panel$), and
$10^{7.5}$ cm$^{-3}$ ($lower$ $panel$) are shown.
In each panel, the models in the range of $10^{20}$ cm$^{-2}$ $\leq$
$N_{\rm H}$ $\leq$ $10^{20}$ cm$^{-2}$ are plotted for the cases of
$U = 10^{-2.5}$, $10^{-2.0}$, and $10^{-1.5}$.
\label{fig3}}
\end{figure*}

In Figure 3, we show the dependences of the flux ratio of
[Fe {\sc vii}]$\lambda$6087/[Ne {\sc v}]$\lambda$3426
on the hydrogen density, the column density and the ionization parameters
of a cloud, and on the presence of dust grains, which are calculated
by our photoionization models with and without dust grains.
The most striking dependence is that on the presence of dust grains.
The models without dust grains predict larger ratios of
[Fe {\sc vii}]$\lambda$6087/[Ne {\sc v}]$\lambda$3426 than the models
with dust grains by a factor of $\sim$10. This factor corresponds to
the depletion factor of iron. This implies that the difference in the
flux ratio between the models with and without grains is mainly due to
the depletion of iron onto dust grains; i.e., the effects of grains
on the thermal equilibrium and on the radiation transfer are negligibly
small. Indeed the models without dust grains but with gas-phase 
elemental abundances of the dusty model predict nearly the same flux 
ratios as those of the dusty models.
This also suggests that the iron depletion is the main reason of 
the difference in the flux ratio of 
[Fe {\sc vii}]$\lambda$6087/[Ne {\sc v}]$\lambda$3426 between the 
models with and without dust grains.

The remaining dependences of the flux ratio are not so large.
As presented in Figure 3, the flux ratio of
[Fe {\sc vii}]$\lambda$6087/[Ne {\sc v}]$\lambda$3426 is almost
independent of the hydrogen density in the range of 
$10^{6.5}$ cm$^{-3}$ $\leq$ $10^{7.5}$ cm$^{-3}$ while it depends on
the hydrogen column density and the ionization parameter.
Note that the models with $U = 10^{-2.5}$ are implausible for gas clouds
in HINERs since those models predict significantly smaller ratios of
[Fe {\sc vii}]$\lambda$6087/[O {\sc iii}]$\lambda$5007 and
[Ne {\sc v}]$\lambda$3426/[O {\sc iii}]$\lambda$5007 
($< 0.01$ and $< 0.1$, respectively)
than the observed values (see, e.g., Murayama \& Taniguchi 1998a;
Nagao et al. 2000b, 2001b).
Thus the dependence of the flux ratio of
[Fe {\sc vii}]$\lambda$6087/[Ne {\sc v}]$\lambda$3426 on the 
hydrogen column density and the ionization parameter in the ranges of 
$10^{20.0}$ cm$^{-2}$ $\leq$ $N_{\rm H}$ $\leq$ $10^{22.0}$ cm$^{-2}$
and $10^{-2.0} \leq U \leq 10^{-1.5}$ is not so large,
i.e., a factor of 2 - 5.
This is consistent with the remark by Ferguson et al. (1997b)
and thus the exact determination of the gas-phase iron abundance
seems to be difficult by simply using the flux ratio of
[Fe {\sc vii}]$\lambda$6087/[Ne {\sc v}]$\lambda$3426.
This is more clearly shown in Figure 4, in which the volume emissivities
of the [Fe {\sc vii}]$\lambda$6087 and [Ne {\sc v}]$\lambda$3426
emission are plotted as functions of the depth into the nebula for 
the models with $n_{\rm H} = 10^{7.0}$ cm$^{-3}$, with
$U = 10^{-2.5}, 10^{-2.0}, 10^{-1.5}$, and with/without dust grains.
The ratio of the emissivities for the two emission lines apparently
depends on the ionization parameter.

Although the flux ratio of 
[Fe {\sc vii}]$\lambda$6087/[Ne {\sc v}]$\lambda$3426 is inappropriate
for the exact determination of the gas-phase iron abundance, it can
be used to discuss the dust abundances in HINERs.
This is because the ranges of the predicted ratio of 
[Fe {\sc vii}]$\lambda$6087/[Ne {\sc v}]$\lambda$3426 are well separated
between the models with and without dust grains.
The dust-free models predict $0.05 \lesssim$
$F$([Fe {\sc vii}]$\lambda$6087)/$F$([Ne {\sc v}]$\lambda$3426) 
$\lesssim 0.3$ while the dusty models predict $0.007 \lesssim$
$F$([Fe {\sc vii}]$\lambda$6087)/$F$([Ne {\sc v}]$\lambda$3426) 
$\lesssim 0.04$. Here we should recall that the models with $U = 10^{-2.5}$
are not taken into account since they are not plausible models for
clouds in HINERs. Using this flux ratio, we discuss the dust 
abundances at HINERs in the next section.

\section{DISCUSSION}

\begin{figure*}
\epsscale{1.80}
\plotone{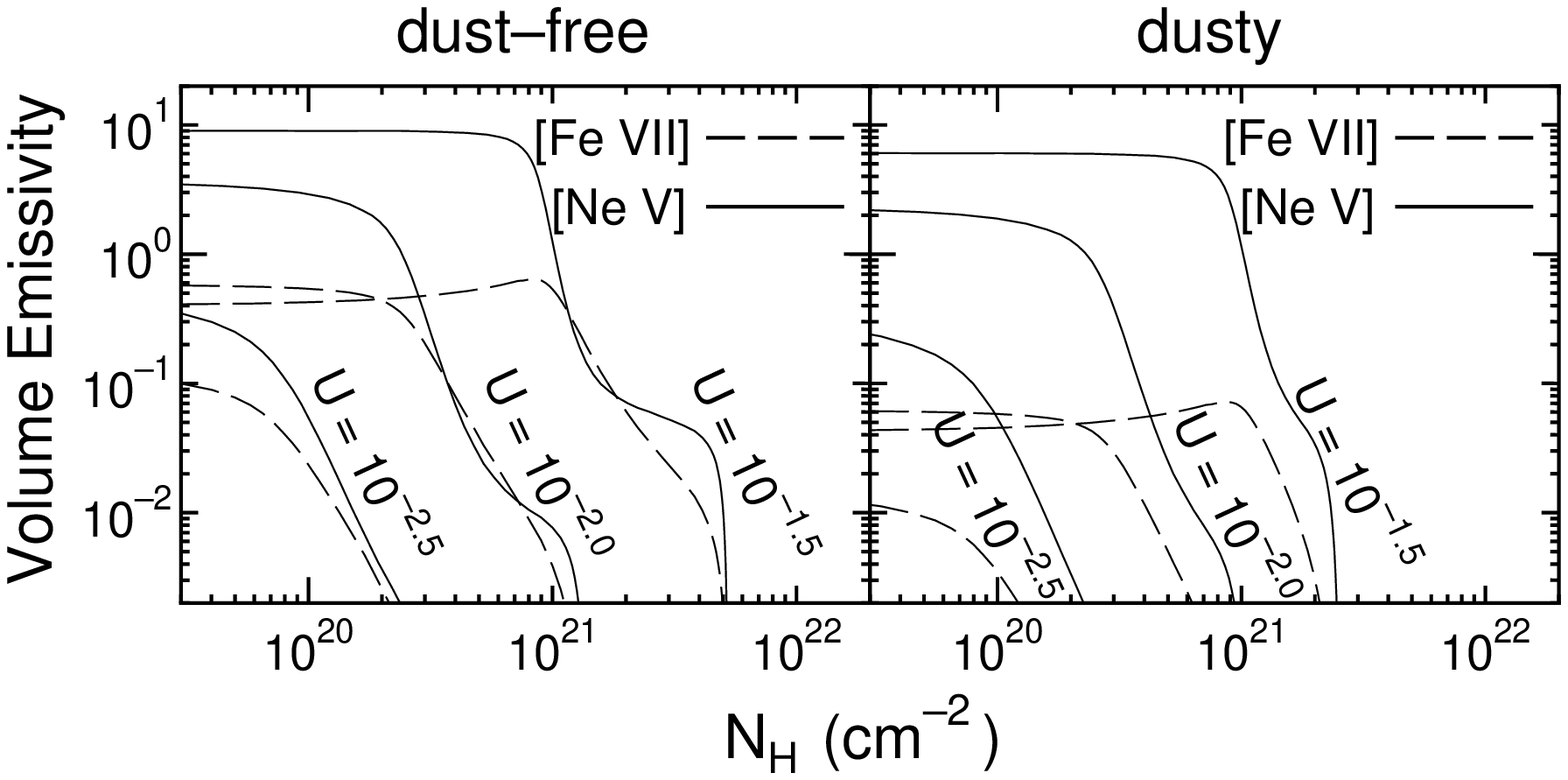}
\caption{
Volume emissivities of [Ne {\sc v}]$\lambda$3426 and 
[Fe {\sc vii}]$\lambda$6087 as functions of depth into the nebula 
for the dust-free models ($left$) and dusty models ($right$) 
with $n_{\rm H} = 10^{7.0}$ cm$^{-3}$ and
$U = 10^{-2.5}$, $10^{2.0}$, and $10^{-1.5}$.
These volume emissivities are normalized
by the H$\beta$ volume emissivity at the hydrogen fully ionized region,
in which the H$\beta$ volume emissivity does not vary significantly.
\label{fig4}}
\end{figure*}

Now we discuss our main problem; are HINERs dusty or not?
The predicted range of the flux ratio of 
[Fe {\sc vii}]$\lambda$6087/[Ne {\sc v}]$\lambda$3426 by both the dust-free
and the dusty models is smaller than the observed range.
This is because the collected data are not corrected for the extinction.
Since a wavelength difference between the two emission lines, 
[Fe {\sc vii}]$\lambda$6087 and [Ne {\sc v}]$\lambda$3426, is large,
the effects of extinction correction are not negligible.
Adopting the extinction curve described by Cardelli, Clayton, \& 
Mathis (1989), and $R_V = A_V / E({\rm B - V}) = 3.1$, 
the correction factor for the dust extinction is 1.94 in the case of
$A_V = 1$ mag. Therefore, the predicted ranges of the flux ratio of 
[Fe {\sc vii}]$\lambda$6087/[Ne {\sc v}]$\lambda$3426 when the
extinction correction of 0 mag $\leq A_V \leq$ 1 mag is taken into account
are $0.05 \lesssim$
$F$([Fe {\sc vii}]$\lambda$6087)/$F$([Ne {\sc v}]$\lambda$3426) 
$\lesssim 0.6$ and $0.007 \lesssim$
$F$([Fe {\sc vii}]$\lambda$6087)/$F$([Ne {\sc v}]$\lambda$3426) 
$\lesssim 0.08$ for the dust-free models and the dusty models,
respectively. Note that the observed Balmer decrements suggest
the reddening amount of 0 mag $\lesssim A_V \lesssim$ 1 mag for type 1 AGNs
(e.g., Cohen 1983; Murayama 1995; Rodr\'{\i}guez-Ardila, Pastoriza, \& 
Donzelli 2000).
The range of the flux ratio predicted by the dust-free models is
roughly consistent with the observed data while that predicted
by the dusty models is far smaller than the observed data.

To explain the observed range of the flux ratio of
[Fe {\sc vii}]$\lambda$6087/[Ne {\sc v}]$\lambda$3426 by the dusty models,
dust extinction of 3 mag $\lesssim A_V \lesssim$ 10 mag is required.
This required range of the dust extinction is far larger than that
estimated from the observed Balmer decrements.
This discrepancy may be more significant if the iron depletion factor
is smaller than the value we adopt here (= 0.1).
In cold ISM, the iron depletion factor reaches down to 0.01
(e.g., Jenkins, Savage, \& Spitzer 1986; Cowie \& Songaila 1986).
Although this large discrepancy might be explained by introducing dust grains
which are located in/around HINERs selectively, this idea has the 
following two serious problems.
First, to assume the existence of additional dust grains in the inner
part of NLRs may conflict with the idea that it is relatively
difficult for the dust grains to survive under high photoionizing flux,
although we cannot exclude the possibility that the grains
could survive under the high flux for some situations.
We should recall the histograms presented in Figure 1, which suggest
that the HINERs are located at the innermost region in NLRs.
Second, the inferred amount of extinction, 
3 mag $\lesssim A_V \lesssim$ 10 mag, corresponds to 
$N_{\rm H^0} \sim (5 - 15) \times 10^{21}$ cm$^{-2}$.
This is far larger than the column density toward the nuclei of 
type 1 AGNs, derived from X-ray spectral analysis ($N_{\rm H^0} 
\lesssim 10^{21}$ cm$^{-2}$; e.g., Reynolds 1997; Leighly 1999).
These considerations suggest that the observed data are hard to be
explained by the dusty models even if very large amount of dust
shielding only the HINERs selectively is introduced.
Therefore, we conclude that the iron in the HINERs is not depleted
onto dust grains significantly.
This conclusion is consistent with the idea that 
HINERs are located at the innermost of NLRs and thus are hidden
by dusty tori when seen from a edge-on view toward dusty tori, which
is implied by the histograms presented in Figure 1
(see also Pier \& Voit 1995; Murayama \& Taniguchi 1998a, 1998b;
Barth et al. 1999; Tran et al. 2000; Nagao et al. 2000b, 2001a, 2001b).
Considering that the HINERs are located closer than the dust-sublimation
radius (i.e., inner radius of dusty tori), we can understand 
the absence of internal dust grains in HINERs and the AGN-type
dependence of the visibility of high-ionization emission lines
as presented in Figure 1, simultaneously.

Finally, we mention the possible difference in the flux ratio of
[Fe {\sc vii}]$\lambda$6087/[Ne {\sc v}]$\lambda$3426 between type 1
and type 2 AGNs. Although the difference is not significant as presented
in Figure 2, the difference in the extinction-corrected flux ratio may
be larger than that in the observed flux ratio since the dust extinction
is larger on average in type 2 AGNs than type 1 AGNs (e.g.,
Dahari \& De Robertis 1988). 
This difference may suggest that iron in spatially extended HINERs
is depleted onto dust grains. Such spatially extended HINERs
have been predicted theoretically (e.g., Korista \& Ferland 1989; 
Ferguson et al. 1997b), emissivities of high-ionization emission lines
at the extended HINERs are expected to be low. Indeed 
such extended HINERs are observationally detected only in few Seyfert galaxies
(e.g., Golev et al. 1995; Murayama et al. 1998; Nagao et al. 2000a;
Nelson et al. 2000; Kraemer \& Crenshaw 2000).
Therefore, the effect of the presence of the spatially extended HINERs
on the observed flux ratios of
[Fe {\sc vii}]$\lambda$6087/[Ne {\sc v}]$\lambda$3426 seems to be low.
However, as suggested by the histograms presented in Figure 1,
the innermost region in NLRs, where strong [Fe {\sc vii}]$\lambda$6087
and [Ne {\sc v}]$\lambda$3426 arise, is hidden as for type 2 AGNs.
In this case, the effect of dusty and spatially extended HINERs
may emerge in the observed flux ratios of 
[Fe {\sc vii}]$\lambda$6087/[Ne {\sc v}]$\lambda$3426.
This idea is a highly speculative one since the observed difference
in the flux ratio of [Fe {\sc vii}]$\lambda$6087/[Ne {\sc v}]$\lambda$3426
between the type 1 and the type 2 AGNs is marginal.
In order to examine this idea observationally, spatial distributions
of the flux ratio of [Fe {\sc vii}]$\lambda$6087/[Ne {\sc v}]$\lambda$3426
should be investigated. Such studies can make it clear how the iron
depletion onto dust grains depends on the location in NLRs.

\section{CONCLUDING REMARKS}

In this paper, we have compared the observed data on two high-ionization
forbidden emission lines, [Fe {\sc vii}]$\lambda$6087 and 
[Ne {\sc v}]$\lambda$3426, with the results of photoionization model
calculations, and found that the HINERs in NLRs of AGNs are not dusty.
This finding was already noted by, e.g., Nussbaumer \& Osterbrock (1970) 
and Ferguson et al. (1997b). However, our finding is important 
because we have shown the absence of dust grains in HINERs
for a large sample of AGNs, for the first time.

In order to examine this conclusion more directly,
we should carry out spatially resolved imaging observation of
the thermal emission of dust in NLRs of AGNs.
By using ALMA (Atacama Large Millimeter/submillimeter Array),
we can resolve structures in the 0.01 arcsec scale by submillimetric
imaging observation. This corresponds to $\sim$0.02 pc for the Circinus
galaxy (adopting the distance of 3.8 Mpc; Freeman et al. 1977).
Thanks to this ability of high spatial resolution, we will be able to
investigate the distribution of dust in NLRs directly.

\acknowledgments

We would like to thank Gary Ferland for providing his code $Cloudy$ 
to the public. We also thank the anonymous referee for useful comments.
A part of this work was financially supported by 
Grants-in-Aid for the Scientific Research (10044052, 10304013, and 
13740122) of the Japanese
Ministry of Education, Culture, Sports, Science, and Technology.


\clearpage

\begin{deluxetable}{lccl}
\scriptsize
\tablenum{1}
\tablecaption{Compiled Data}
\tablehead{
\colhead{Object} &
\colhead{[Ne {\sc v}]/[O {\sc iii}]} &
\colhead{[Fe {\sc vii}]/[O {\sc iii}]} &
\colhead{References\tablenotemark{a}}
}
\startdata  
\cutinhead{Type 1 AGN}
NGC 3227 & 0.035 & 0.007 & 1 \nl
NGC 3783 & 0.107 & 0.103 & 2, 3, 4 \nl
NGC 4051 & 0.218 & 0.041 & 3, 5, 6, 7 \nl
NGC 4593 & 0.146 & 0.178 & 2 \nl
NGC 5548 & 0.179 & 0.060 & 1, 3, 7, 8 \nl
NGC 7469 & 0.151 & 0.044 & 1, 2, 3, 7, 8, 9, 10\nl
Mrk 9    & 0.371 & 0.202 & 5, 10 \nl
Mrk 42   & 0.258 & 0.060 & 6, 11 \nl
Mrk 79   & 0.107 & 0.027 & 1, 3, 8 \nl
Mrk 335  & 0.295 & 0.122 & 3, 10, 12 \nl
Mrk 359  & 0.200 & 0.093 & 5, 11, 13 \nl
Mrk 376  & 0.218 & 0.127 & 10 \nl
Mrk 699  & 0.286 & 0.116 & 5, 14, 15, 16, 17 \nl
Mrk 704  & 0.234 & 0.115 & 1, 5, 18 \nl
Mrk 766  & 0.048 & 0.081 & 7, 11 \nl
Mrk 783  & 0.029 & 0.012 & 11 \nl
Mrk 817  & 0.334 & 0.020 & 1, 12 \nl
Mrk 841  & 0.201 & 0.031 & 1, 2 \nl
Mrk 871  & 0.219 & 0.074 & 2, 5 \nl
Mrk 926  & 0.124 & 0.009 & 1, 4, 19 \nl
Mrk 975  & 0.187 & 0.073 & 1, 5 \nl
Mrk 1126 & 0.109 & 0.067 & 11 \nl
Mrk 1393 & 0.076 & 0.023 & 2 \nl
Akn 120  & 0.217 & 0.242 & 5, 10 \nl
CTS F10.01  &0.258&0.100 & 20 \nl
ESO 141-G55 &0.250&0.077 & 2, 3, 4 \nl
Fairall 9   &0.220&0.089 & 3, 4 \nl
Fairall 51  &0.219&0.070 & 2 \nl
Fairall 1116&0.590&0.390 & 4 \nl
MCG -6-30-15&0.092&0.208 & 2 \nl
MCG  8-11-11&0.047&0.038 & 1, 3 \nl
UGC 11763& 0.300 & 0.072 & 8, 10 \nl
3C 120   & 0.104 & 0.016 & 3, 8, 19 \nl
3C 445   & 0.077 & 0.038 & 2, 5 \nl
\cutinhead{Type 2 AGN}
NGC 424  & 0.126 & 0.054 & 3, 21 \nl
NGC 1068 & 0.054 & 0.012 & 3, 22, 23 \nl
NGC 1386 & 0.063 & 0.035 & 24 \nl
NGC 3281 & 0.063 & 0.018 & 19, 24 \nl
NGC 4507 & 0.063 & 0.018 & 3, 24 \nl
NGC 5506 & 0.025 & 0.005 & 2, 3, 6, 25 \nl
NGC 5728 & 0.030 & 0.052 & 24 \nl
NGC 6890 & 0.211 & 0.056 & 24 \nl
NGC 7674 & 0.066 & 0.016 & 6, 7, 9, 23, 26 \nl
Mrk 1    & 0.047 & 0.019 & 6, 9, 22 \nl
Mrk 34   & 0.044 & 0.007 & 6, 22 \nl
Mrk 78   & 0.023 & 0.010 & 6, 22 \nl
Mrk 348  & 0.055 & 0.017 & 6, 9, 22 \nl
Mrk 463E & 0.017 & 0.003 & 6, 26 \nl
Mrk 477  & 0.056 & 0.008 & 15, 17, 26 \nl
Akn 347  & 0.068 & 0.014 & 26 \nl
IC 5135  & 0.120 & 0.020 & 24 \nl
PKS 2048-57&0.018& 0.007 & 3, 24 \nl 
Tololo 1351-375&0.067&0.012& 2 \nl
UM 16    & 0.057 & 0.017 & 6, 26 \nl
3C 33    & 0.041 & 0.011 & 22 \nl
3C 223   & 0.066 & 0.011 & 27 \nl
3C 223.1 & 0.110 & 0.028 & 27 \nl
3C 327   & 0.084 & 0.023 & 3, 28 \\
\enddata 
\tablenotetext{a}{References. ---
                  (1)  Cohen 1983;
                  (2)  Morris \& Ward 1988;
                  (3)  Penston et al. 1984;
                  (4)  Winkler 1992;
                  (5)  Erkens, Appenzeller, \& Wagner 1997;
                  (6)  Malkan 1986;
                  (7)  Murayama 1995;
                  (8)  Osterbrock 1977;
                  (9)  Cruz-Gonz\'{a}lez et al. 1994;
                  (10) Phillips 1978;
                  (11) Osterbrock \& Pogge 1985;
                  (12) Grandi 1983;
                  (13) Davidson \& Kinman 1978;
                  (14) Ferland \& Osterbrock 1987;
                  (15) Kunth \& Sargent 1979;
                  (16) Osterbrock 1981;
                  (17) O'Connell \& Kingham 1978;
                  (18) Ulvestad \& Wilson 1983;
                  (19) Durret \& Bergeron 1988;
                  (20) Rodr\'{\i}guez-Ardila, Pastoriza, \& Maza 1998;
                  (21) Fosbury \& Sansom 1983;
                  (22) Koski 1978;
                  (23) Osterbrock \& Martel 1993;
                  (24) Phillips, Charles, \& Baldwin 1983;
                  (25) Shuder 1980;
                  (26) Shuder \& Osterbrock 1981;
                  (27) Cohen \& Osterbrock 1981;
                  (28) Costero \& Osterbrock 1977.
}
\end{deluxetable}

\end{document}